\pdfoutput=1
\documentclass[11pt,twoside,a4paper,cmspaper,final,collab]{cms-tdr}
\def\svnVersion{829d365}\def\svnDate{2026/04/09}\def\cmsCernNoTag{CERN-EP-2025-095}\def\cmsCernDate{\today}\def\cmsMessage{Published in Physical Review Letters as \href{http://dx.doi.org/10.1103/lckg-sdh9}{\doi{10.1103/lckg-sdh9}.}}
\begin{document}\cmsNoteHeader{HIN-24-003}

\hyphenation{had-ron-i-za-tion}
\hyphenation{cal-or-i-me-ter}
\hyphenation{de-vices}

\newcommand{\Qsquared}{\ensuremath{Q^{2}}\xspace}
\newcommand {\PbPb}  {\mbox{PbPb}}
\newcommand {\sNN}  {\ensuremath{s_{_{\mathrm{NN}}}}\xspace}
\ifthenelse{\boolean{cms@external}}
{
\newcommand{\Xn}{\ensuremath{\mathrm{X}\mkern-2mu\Pn}\xspace}
\newcommand{\Zn}{\ensuremath{0\Pn}\xspace}
\newcommand {\XnZeron} {\ensuremath{\Xn\Zn}\xspace}
\newcommand {\ZeronXn} {\ensuremath{\Zn\mkern-2mu\Xn}\xspace}
}
{
\newcommand{\Xn}{\ensuremath{\mathrm{X}\Pn}\xspace}
\newcommand{\Zn}{\ensuremath{0\Pn}\xspace}
\newcommand {\XnZeron} {\ensuremath{\Xn\Zn}\xspace}
\newcommand {\ZeronXn} {\ensuremath{\Zn\Xn}\xspace}
}

\cmsNoteHeader{HIN-24-003}
\title{Measurement of \texorpdfstring{$\PDz$}{D0} meson photoproduction in ultraperipheral heavy ion collisions}

\date{\today}

\abstract{
This Letter reports the first measurement of photonuclear \PDz meson production in ultraperipheral heavy ion collisions. The study is performed using lead-lead collision data, with an integrated luminosity of 1.34\nbinv, collected by the CMS experiment at a nucleon-nucleon center-of-mass energy of 5.36\TeV. Photonuclear events, where one of the colliding nuclei breaks up and the other remains intact, are selected based on breakup neutron emissions and by requiring no particle activity in a large rapidity interval in the direction of the photon-emitting nucleus. The \PDz mesons are reconstructed via the $\PDz\to\PKm\PGpp$ decay channel, with the cross section measured as a function of \PDz meson transverse momentum and rapidity. The results are compared with next-to-leading-order perturbative QCD calculations that employ recent parametrizations of the lead nuclear parton distribution functions, as well as with predictions based on the color glass condensate framework. This measurement is the first photonuclear collision study characterizing parton distribution functions of lead nuclei for parton fractional momenta $x$ (relative to the nucleon) ranging approximately from a few $10^{-4}$ to $10^{-2}$ for different hard energy scale \Qsquared selections. 
\par\vspace*{2ex}
\textit{This work is dedicated to the memory of our colleague and friend, Michele Arneodo. Michele was a generous mentor, a passionate physicist, and a kind soul who inspired both students and collaborators with his wisdom, warmth, and dedication to science. His contributions to diffractive physics and his support for young scientists leave a lasting legacy. He will be deeply missed}
}
\hypersetup{
pdfauthor={CMS Collaboration},
pdftitle={Measurement of D0 meson photoproduction in ultraperipheral heavy ion collisions},
pdfsubject={CMS},
pdfkeywords={CMS, UPC 0nXn, gluon saturation, charm photo-production}}

\maketitle 
Ultraperipheral collisions (UPCs) of heavy ions serve as a powerful experimental tool for studying the gluonic structure of nuclei and for probing the evolution equations of quantum chromodynamics (QCD) for partons carrying a small fraction of the nucleon momentum, $x$~\cite{Strikman:2005yv, Klein:2020fmr}. The UPCs occur when the impact parameter of the collision exceeds the sum of the radii of the two nuclei. At ultrarelativistic energies, the Lorentz-contracted electromagnetic fields surrounding the heavy ions act as sources of high-energy quasireal photons, leading to abundant photonuclear interactions. An overview of recent UPC results from CMS can be found in Ref.~\cite{CMS:2024krd}. 

For LHC energies in particular, the large flux of high-energy photons results in sizeable cross sections for the production of heavy quarks~\cite{Klein:2002wm,Cacciari:2025tgr}. Charm photoproduction in UPCs occurs when a quasireal photon emitted from one nucleus interacts with a gluon from the other nucleus, forming a charm quark-antiquark pair (\ccbar). Measurements of the production yields of charmed hadrons, which result from the hadronization of \PQc quarks, are sensitive to the gluon distribution in the target nucleus~\cite{Behnke:2015qja}. Because of their large masses relative to $\Lambda_\text{ QCD}$, the energy scale below which strong coupling becomes large, the production of heavy quarks can be described by perturbative QCD (pQCD) calculations down to zero transverse momentum \pt. In photonuclear events, the energy scale of the interaction (\Qsquared) and the $x$ value of the scattered partons in the nucleus can be inferred from the transverse momenta and rapidities of the final-state \PQc quarks. Consequently, by measuring charmed meson yields as functions of \pt and rapidity ($y$) the properties of gluons in the nucleus can be constrained across different $x$ and \Qsquared regions. Similarly, jets produced in UPCs have been used to probe the properties of partons in the nucleus~\cite{ATLAS:2024mvt}. The low particle multiplicities and negligible backgrounds from strong interactions that characterize such events ensure that the fragmentation and hadronization of \PQc quarks in UPCs occur in a vacuum-like environment, without significant modifications caused by the presence of a large number of color-carrying partons~\cite{ALICE:2016yta,LHCb:2017yua,CMS:2018loe,ALICE:2020wfu,LHCb:2023rpm}. Calculations of heavy-quark production in UPCs have so far been performed at leading order in perturbation theory either within the collinear factorization approach~\cite{Klein:2002wm,Goncalves:2001vs,Goncalves:2003is,Adeluyi:2012ds} or using the color glass condensate (CGC) framework~\cite{Goncalves:2003is,Gimeno-Estivill:2025rbw}. More recently, G$\gamma$A-FONLL~\cite{Cacciari:2025tgr} has extended these studies to next-to-leading-order accuracy within collinear factorization.

In this Letter, we present the first measurement of the photonuclear production cross section of inclusive \PDz mesons (including beauty-hadron decays) in UPCs. The data presented here probe, for the first time in photonuclear collisions, the parton distribution functions of lead nuclei for $x$ ranging from about $3\ten{-4}$ to $3\ten{-2}$ and \Qsquared from about 18 to $600\GeV^{2}$. Tabulated results are provided in the \textsc{HEPData} record for this analysis~\cite{HEPData}.

The central feature of the CMS apparatus is a superconducting solenoid of 6\unit{m} internal diameter, providing a magnetic field of 3.8\unit{T}. Within the solenoid volume are a silicon pixel and strip tracker, a lead tungstate crystal electromagnetic calorimeter, and a brass and scintillator hadron calorimeter, each composed of a barrel and two endcap sections. Forward calorimeters (HF) extend the pseudorapidity ($\eta$) coverage provided by the barrel and endcap detectors. The zero degree calorimeters (ZDCs) are used to measure the energy of very forward neutrons produced from nuclear breakup~\cite{Suranyi:2021ssd,CMS-DP-2024-002}. The ZDCs are located in front of the neutral particle absorber, roughly ${\pm}140\unit{m}$ away from the CMS interaction point, between the two beampipes. They measure neutral particles at $\abs{\eta}> 8.5$. Particles are reconstructed using the CMS particle-flow algorithm~\cite{CMS:2017yfk} which combines the information of the different subdetectors to determine the kinematics of each event. The CMS experiment uses a two-tiered trigger system~\cite{Khachatryan:2016bia}. The first level (\Lone or L1) comprises custom hardware processors and uses fast information from the calorimeters and the muon system to achieve the first significant data rate reduction. The second level, \ie, the high-level trigger, implements a simplified version of the CMS offline reconstruction software running on a computer farm and can perform more refined selections.  A more detailed description of the CMS detector, together with a definition of the coordinate system used and the relevant kinematic variables, can be found in Ref.~\cite{CMS:2008xjf}. Details of the CMS detector configuration at the time of this measurement can be found in Ref.~\cite{CMS:2023gfb}.

The analysis uses the lead-lead (\PbPb) data set, corresponding to a recorded integrated luminosity of about 1.34\nbinv, collected in 2023 at a nucleon-nucleon center-of-mass energy of 5.36\TeV. The photonuclear production of \PDz mesons in the \pt range 2--5\GeV is measured using a sample of events triggered by a signal in at least one of the ZDCs that is greater than or equal to the 1-neutron (1\Pn) threshold (inclusively referred to as \Xn). A dedicated L1 trigger algorithm requiring an \Xn signal in one ZDC and the absence of signal in the other (0\Pn), in coincidence with an L1 jet with energy above 8\GeV, was used to enrich the sample of \PDz mesons with $\pt>5\GeV$. In this Letter, the notation \XnZeron (\ZeronXn) indicates that the outgoing lead-ion at negative (positive) rapidities breaks up. Offline, events are required to have one primary vertex, formed by at least two tracks and located within 15\cm of the nominal interaction region along the beamline. Events must additionally fulfill selection criteria intended to reject beam-gas interactions and accelerator-induced backgrounds~\cite{CMS:2019ctu}. To suppress residual contamination from hadronic events, a large rapidity gap is required on the side of the outgoing intact nucleus which emitted the photon, following previous analyses~\cite{ATLAS:2021jhn,ATLAS:2024mvt,ATLAS:2025tof}. This condition is satisfied if no particle-flow candidate with energy above the noise threshold is detected in the $\eta$ range of the HF calorimeter ($3.0 < \abs{\eta} < 5.2$). An energy threshold of 9.2 (8.6)\GeV is applied to the HF detector at positive (negative) $\eta$, with values optimized according to each detector's performance. Monte Carlo (MC) simulations are used to estimate signal efficiency, detector acceptance, and some background sources. Photonuclear events are generated with \PYTHIA version 8.309~\cite{Bierlich:2022pfr}, with default photon flux settings for Pb ions acting as a source of quasireal photons in the equivalent photon approximation~\cite{Helenius:2018bai}. The events include direct and resolved photoproduction processes, with the photon PDF modeled using the Cornet-Jankowski-Krawczyk-Lorca, or CJKL, parametrization~\cite{PhysRevD.68.014010}, and the lead nuclear PDF (nPDF) parametrized by EPPS21~\cite{Eskola:2021nhw}. Both promptly ($\PQc\to\PDz$) and nonpromptly ($\PQb\to\PDz$) produced \PDz mesons are included in the simulation. The events are propagated through the CMS detector with \GEANTfour~\cite{geant4}, with \PDz meson decays simulated using \EVTGEN2.0.0~\cite{Lange:2001uf}, and final-state photon radiation modeled by \PHOTOS 2.0~\cite{Barberio:1990ms}.

\begin{figure}[ht!]
    \centering
    \includegraphics[width=0.45\textwidth]{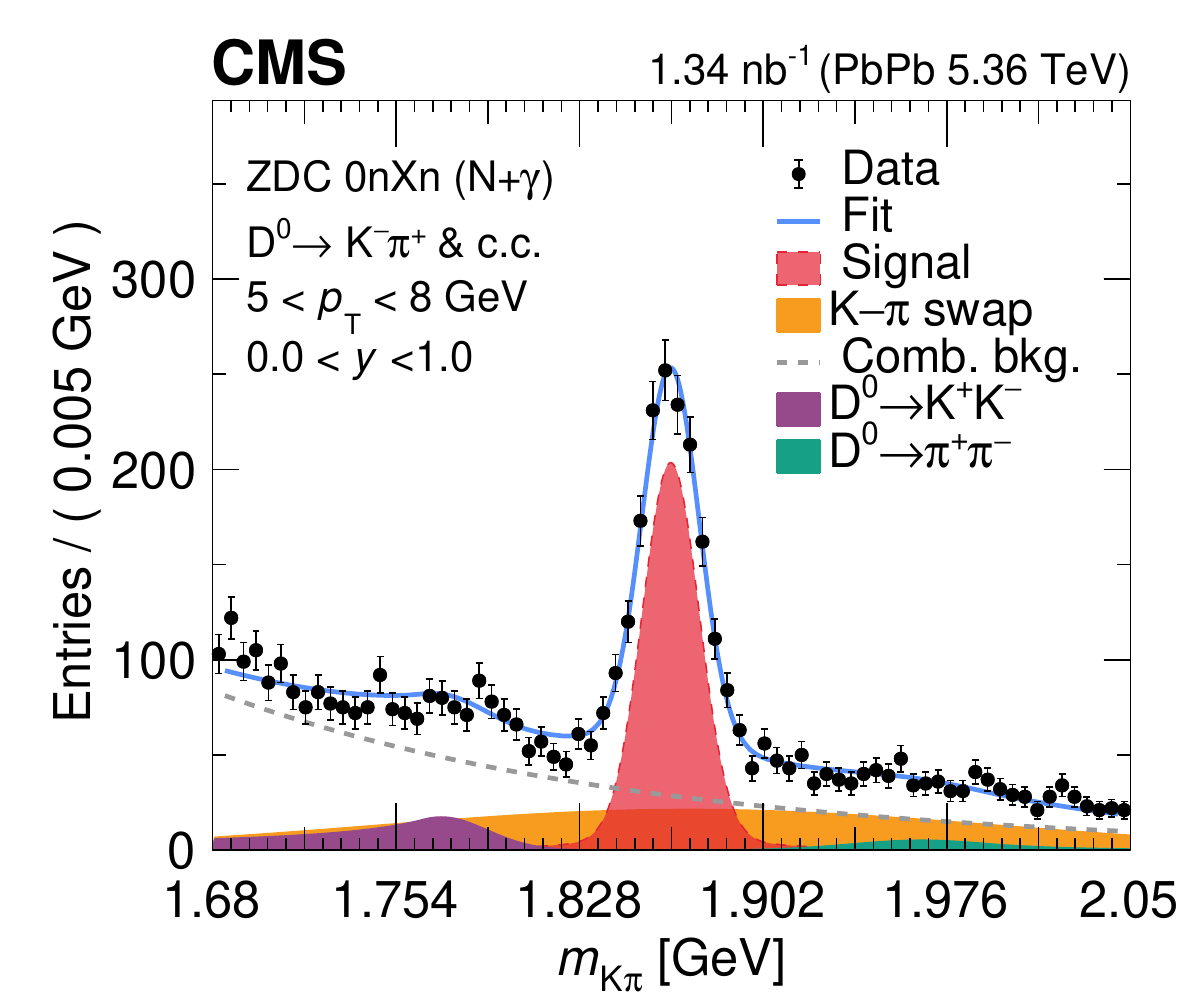}
    \caption{Invariant mass distribution of \PDz mesons with $5 < \pt < 8\GeV$ and $0.0 < y < 1.0$. The description of the fit template is provided in the text.}
    \label{fig:massFit}
\end{figure}

The \PDz mesons and their charge conjugates (\PADz mesons) are reconstructed via the hadronic decay channel $\PDz\to\PKm\Pgpp$, following a strategy similar to that discussed in Ref.~\cite{CMS:2017qjw}. Only \PDz candidates with daughter tracks satisfying high-purity selection criteria~\cite{Chatrchyan:2014fea}, along with the conditions $\pt > 1\GeV$ and $\abs{\eta}<2.4$ are considered. The \PDz candidates are then reconstructed by combining pairs of oppositely charged tracks with an invariant mass within 185\MeV of the world-average \PDz meson mass of 1868.48\MeV~\cite{ParticleDataGroup:2024cfk}. This range provides a handle to constrain the modeling of the backgrounds described later in this Letter. For each track pair, two \PDz candidates are created by assigning pion and kaon masses to the tracks in both possible combinations. To reduce the combinatorial background, the \PDz meson candidates are selected based on the value of four topological variables (where the selection value used depends on the \pt and $y$ of the candidate): the three-dimensional decay length normalized to its uncertainty (required to be larger than 2.5--3.5), the pointing angle defined as the angle between the total momentum vector of the tracks and the vector connecting the primary and the secondary vertices (required to be smaller than 0.25--0.50\unit{rad}), the opening angle between the two daughter tracks (required to be smaller than 0.3--0.5\unit{rad}), and the $\chi^2$ probability of the \PDz vertex fit~\cite{Forden:161017} (p-value required to be larger than 0.1). The selection is optimized in each \pt bin to maximize the expected statistical significance of the \PDz meson signal. The \PDz meson yields are extracted in three different \pt intervals (2--5, 5--8, and 8--12\GeV). Four rapidity intervals ($-2 < y < -1$, $-1 < y < 0$, $0 < y < 1$, and $1 < y < 2$) are used for \PDz mesons in bins with $\pt >5\GeV$, whereas a single rapidity interval ($-1 < y < 1$) is used for \pt interval 2--5\GeV. The \PDz yields are extracted using an unbinned maximum likelihood fit to the invariant mass distributions in the range $1.68 < m_{\PK\Pgp} < 2.05\GeV$. The yields are determined separately for the \XnZeron and \ZeronXn event categories. In \XnZeron events, the incoming photon-emitting nucleus originates from negative rapidities ($\PGg\mathrm{N}$), whereas in \ZeronXn events it comes from positive rapidities ($\mathrm{N}\PGg$). An example of the \PDz invariant mass distribution is shown in Fig.~\ref{fig:massFit}.

The combinatorial background, arising from track pairs unrelated to genuine \PDz meson decays, is modeled with an exponential function whose slope and normalization are determined from the fit. The signal shape consists of a superposition of two Gaussian functions sharing the same mean but differing in width, a choice made in earlier CMS publications that models the mass peak well~\cite{CMS:2017qjw}. An additional Gaussian function models the invariant mass of \PDz candidates with incorrect mass assignment, for which the mass hypothesis for the pion and kaon are interchanged. The widths of the Gaussian functions for both the \PDz signal and the swapped mass candidates are constrained using MC simulation. A floating multiplicative scaling factor for the width of the distribution is included in the fit to account for potential discrepancies in the signal resolution between data and MC. The ratio of the signal yield to the yield of \PDz candidates with swapped mass assignments is fixed based on MC simulation. A Crystal Ball function is used to describe the peaking backgrounds from the decays $\PDz \to \PKp\PKm$ and $\PDz\to\Pgpp\Pgpm$~\cite{ParticleDataGroup:2024cfk}, with its parameters and the ratio of its integral to the $\PDz\to\PKm\Pgpp$ yield also fixed according to MC simulations. Of the eighteen fits for each \PDz \pt and $y$ bin in \XnZeron and \ZeronXn events, seventeen result in p-values greater than 0.05, and one has p-value of $\sim$0.01. This distribution of p-values is consistent with statistical expectations.

The yields extracted from the fit are corrected for the average trigger prescale factor (applied to reduce the data rate resulting from the high instantaneous luminosity of the LHC), the trigger efficiency, and the efficiency of the offline event selection. The jet trigger efficiency correction is evaluated in intervals of \PDz \pt and rapidity with a two-step procedure. First, the efficiency of the L1 jet selection is evaluated as a function of the leading-track \pt and $\eta$ with a sample of inclusive photonuclear events. The efficiency map is then used to reweight, on an event-by-event basis, the uncorrected distributions of \PDz meson candidates obtained via the triggered sample in intervals of \PDz \pt and $y$. The resulting jet trigger efficiency ranges from 21 to 28\%, depending on the rapidity, for \PDz mesons in the transverse momentum range $5<\pt<8\GeV$, to 46--54\% in the $8<\pt<12\GeV$ range (no jet trigger is used for $2<\pt<5\GeV$). The event selection efficiency, which includes the effect of losses from the rapidity gap and primary vertex requirements, exceeds 98\% across all rapidity and \pt intervals, with comparable values for direct and resolved photoproduction. An additional correction accounts for soft electromagnetic interactions from independent \PbPb~collisions within the same bunch crossing (electromagnetic pileup), which may cause neutron emission from the photon-emitting nucleus, thereby reducing the ZDC \XnZeron (or \ZeronXn) selection efficiency. The correction factor (about 0.96) is estimated from the average amount of pileup interactions in \PbPb~collisions and the cross section for single and double electromagnetic dissociation (EMD)~\cite{ALICE:2012aa,ATLAS:2024mvt}. The acceptance and reconstruction efficiency of \PDz candidates is evaluated as a function of the \PDz \pt and $y$, and ranges from 5\% for $2<\pt<5\GeV$ to 25\% for $8<\pt<12\GeV$.

The measured cross section is subject to several systematic uncertainties arising from the signal extraction, acceptance and reconstruction efficiency, offline selection, trigger selection, branching fraction of the $\PDz\to\PKm\Pgpp$ decay channel, and integrated luminosity determination. The uncertainty in the rapidity-gap selection efficiency is evaluated by varying the energy threshold for the particle-flow candidates used to determine the presence of a gap between 5 and 15\GeV, resulting in an uncertainty of 7--23\%. The correction for the inefficiency induced by electromagnetic pileup is found to have an uncertainty of 4\%. The uncertainty due to the background modeling in the fit is estimated by repeating the fit with different functions for the background. The signal shape is also varied by forcing the means and widths (each separately) of the Gaussian functions that describe the signal to be equal to the values extracted in simulations; these three fit variations result in an uncertainty of 3--20\% depending on the \PDz \pt and $y$. In the background variation study, a second-order Chebyshev polynomial function is used. The uncertainty in the signal yield coming from the modeling of the $\PDz\to\PKp\PKm$ and $\Pgpp\Pgpm$ peaks is estimated from MC to be about 13\%. The final uncertainty in the extracted yield is evaluated as the quadratic sum of all the individual uncertainties. 

Dedicated studies are conducted to estimate the effects of the differences in the properties of signal events in data and simulation. Specifically, to evaluate the impact of the discrepancies in the distributions of the \PDz meson selection variables, the corrected yields obtained by varying each topological selection are studied. Topological selections are relaxed in a range such that signal events can still be extracted with good statistical significance. The final uncertainty
is evaluated as the quadratic sum of the differences between the cross sections
obtained when varying each selection and the nominal one (2--15\%). The difference between the \PDz meson reconstruction and selection efficiencies evaluated in direct- and resolved-photon events is taken to account for the differences in the relative abundance of the two classes of signal events in simulation and data; the resulting effect is 1--5\%. An analogous systematic uncertainty is included to account for differences in the fraction of prompt \PDz mesons ($f_\text{prompt}$) between data and simulation. This uncertainty is evaluated by reweighting the MC-based efficiency for prompt and nonprompt \PDz mesons according to the $f_\text{prompt}$ value extracted from the data. The extraction relies on the difference between prompt and nonprompt events in the distribution of the \PDz meson candidate's distance of closest approach, defined as the decay length multiplied by the sine of the pointing angle. The associated uncertainty is approximately 5\% across all \pt and $y$ intervals.

The \PDz selection and reconstruction efficiency also depends on the distribution of the \PDz mesons in \pt and $y$, as well as on the event multiplicity; this is because the single-track reconstruction efficiency decreases in events with more tracks. The MC samples are reweighted to two alternative distributions, one based on fixed-order-next-to-leading logarithmic (FONLL) calculations and the other on data after accounting for EMD. The charged-hadron multiplicity is also reweighted to match that in data. These variations of the \PDz meson reconstruction and identification efficiency result in a systematic uncertainty of 7\% or less. The uncertainty due to the \PDz trigger efficiency is found to be about 20\% in the \pt bins where the jet trigger is used. The systematic uncertainty in the hadron tracking efficiency (2.3\% per track) is taken from the analysis of the proton-proton collision samples collected in 2022--2023 with comparable detector conditions to those of the current data set~\cite{CMS:2025fmx}. The systematic uncertainty associated with the branching fraction is 0.76\%~\cite{ParticleDataGroup:2024cfk}, and the uncertainty in the integrated luminosity is 6.4\%~\cite{CMS:2025rzq}.

All systematic uncertainties are treated as symmetric. The total systematic uncertainty in the cross section measurement is computed as the sum in quadrature of the different contributions mentioned above, and is found to range from ${\sim}26\text{--}48\%$, depending on the \PDz meson $y$ and \pt. The dominant source of systematic uncertainties in the 2--5\GeV bin is associated with the modeling of $\PDz\to\PKp\PKm$ and $\Pgpp\Pgpm$ decays, the change to a Chebyshev polynomial in modeling the background, and the variation of the pointing angle selection. For 5--8 and 8--12\GeV, the systematic uncertainties are dominated by the trigger and rapidity gap thresholds uncertainties, in addition to the modeling of the $\PDz\to\PKp\PKm$ and $\Pgpp\Pgpm$ decays. Each source of systematic uncertainty is assumed to be fully correlated across bins, whereas the statistical uncertainties are treated as uncorrelated.

\begin{figure*}[tb!]
    \centering
    \includegraphics[width=0.32\textwidth]{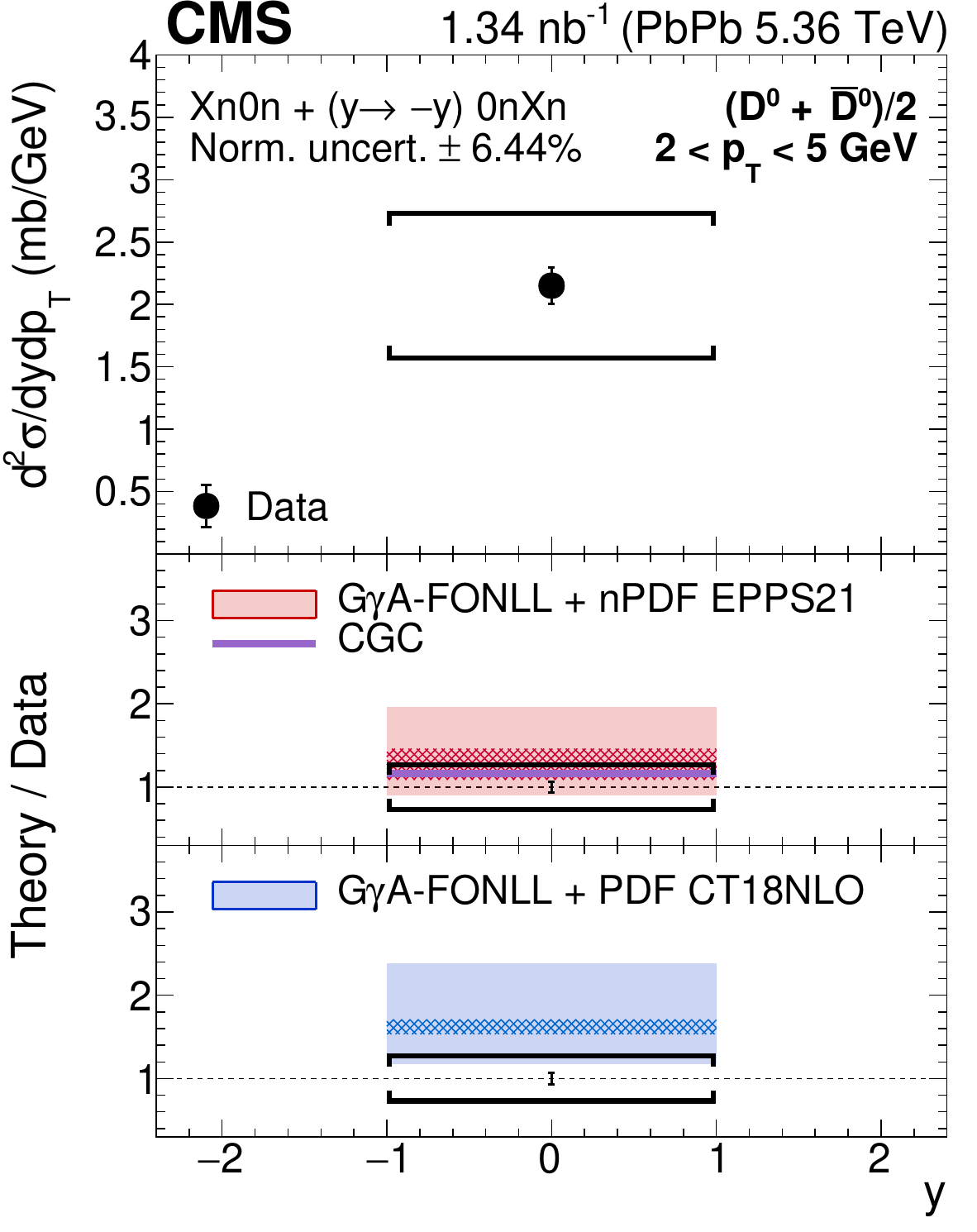}
    \includegraphics[width=0.32\textwidth]{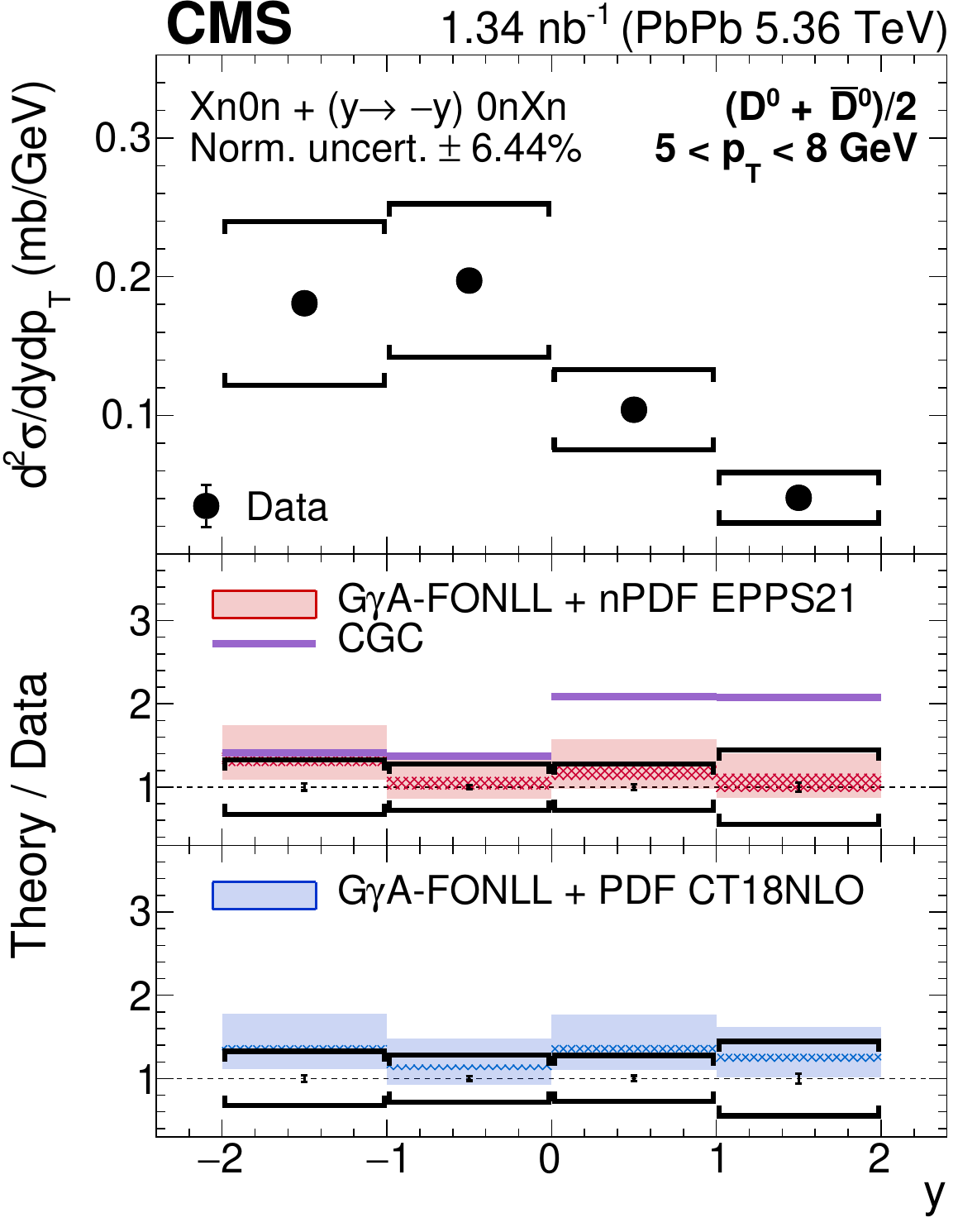}
    \includegraphics[width=0.32\textwidth]{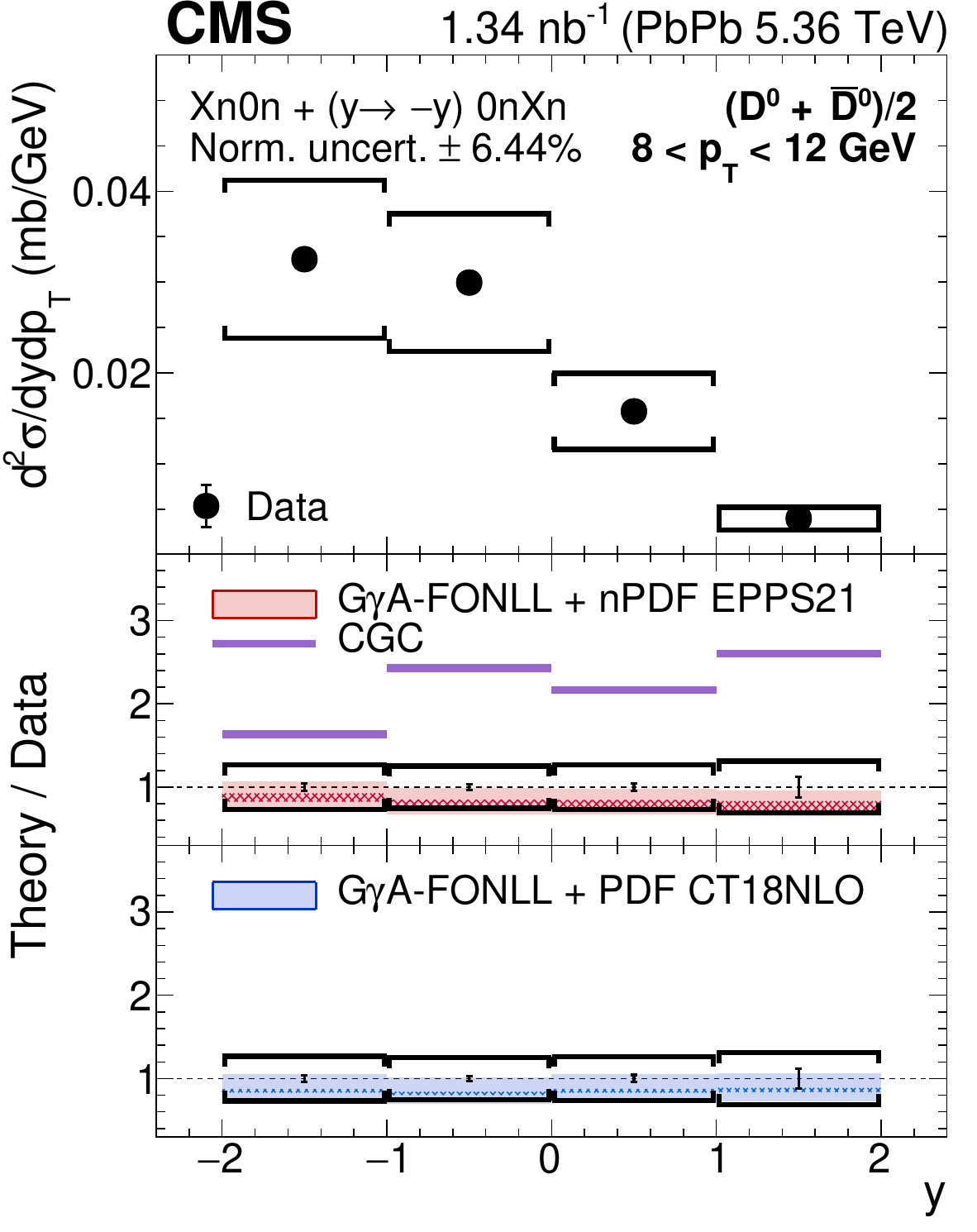}
    \caption{Cross sections for \PDz production in $\PGg\mathrm{N}$ events ($\XnZeron+\ZeronXn$) in three \pt intervals. Vertical bars (brackets) indicate statistical (systematic) uncertainties. Sub-panels show theory-over-data ratios: the middle panel compares G$\gamma$A-FONLL with EPPS21 lead nPDFs and a CGC calculation~\cite{Gimeno-Estivill:2025rbw}, while the bottom panel shows G$\gamma$A-FONLL with CT18NLO proton PDFs. Light-shaded bands reflect scale variations, hatched bands represent PDF and nPDF uncertainties. Vertical bars (brackets) represent statistical (systematic) uncertainties on data.}
    \label{fig:theoryComparisonsEPPS21}
\end{figure*}

In Fig.~\ref{fig:theoryComparisonsEPPS21}, the \PDz meson production cross section in \XnZeron and rapidity-reflected \ZeronXn events is shown (black markers) as a function of the \PDz rapidity in \pt\ intervals. The cross section is divided by a factor of two to average the contributions of \PDz and \PADz mesons. Vertical bars indicate the statistical uncertainties, while brackets represent the systematic ones. Forward (positive) rapidities probe smaller values of $x$, whereas backward (negative) rapidities are sensitive to larger $x$. Likewise, low-\pt\ \PDz mesons correspond to lower \Qsquared values, whereas high-\pt\ hadrons probe the nPDFs at higher \Qsquared. A qualitative estimate of the kinematic region probed in each \PDz meson \pt\ and $y$ interval can be obtained by approximating \Qsquared as $p_{\mathrm{T},c}^{2} + m_{c}^{2}$, and the parton longitudinal momentum fraction in the nucleus as $x \approx \re^{-y}\sqrt{\smash[b]{\Qsquared/\sNN}}$, where \sqrtsNN~is the nucleon-nucleon center-of-mass energy. Assuming the heavy-flavor hadron carries about half of the parent charm-quark momentum, a \PDz meson produced with $\pt = 2\GeV$ at $y = 1$ probes partons with $x\approx3\ten{-4}$ at a hard scale $\Qsquared \approx 18\GeV^{2}$. Conversely, a \PDz meson produced with $\pt = 12\GeV$ at $y = -2$ probes partons with $x \approx 0.03$ at $\Qsquared \approx 600\GeV^{2}$. Further details on the $x$ and \Qsquared coverage of this measurement are given in Ref.~\cite{Cacciari:2025tgr}.

The measurements are compared with recent pQCD calculations from G$\gamma$A-FONLL~\cite{Cacciari:2025tgr}, a framework that builds on FONLL~\cite{Cacciari:1998it,Cacciari:2001td,Frixione:2002zv} to model heavy-quark production in photonuclear collisions and employs photon-flux reweighting to reproduce the flux properties expected in UPCs. The predictions include corrections for the EMD survival probability of the photon-emitting nucleus~\cite{Eskola:2024fhf}. The subpanels of Fig.~\ref{fig:theoryComparisonsEPPS21} display the theory-to-data ratios obtained with lead nPDFs from EPPS21~\cite{Eskola:2021nhw} (middle) and proton PDFs from CT18NLO~\cite{Hou:2019efy} (bottom).
The light blue and red bands represent the uncertainties associated with variations of the FONLL factorization and renormalization scales, while the hatched bands represent the nPDF-parametri\-zation uncertainties. For \PDz mesons with $2<\pt<5\GeV$, the theory-to-data ratios obtained with EPPS21 lie slightly above unity (about 1.4). Although still compatible within the combined experimental and theoretical uncertainties, this trend may indicate a stronger nuclear suppression for low-$x$ gluons than predicted. In the highest-\pt\ interval, $8<\pt<12\GeV$, the theory-to-data ratio obtained with EPPS21 remains consistently below unity, at about 0.8--0.9, across the probed \PDz meson rapidity range, indicating that the data slightly exceed the nPDF-based predictions. In the same subpanel, the data are also compared with recent theoretical predictions~\cite{Gimeno-Estivill:2025rbw} based on the CGC framework, which relies on nonlinear QCD evolution to model the properties of gluons at low~$x$. The CGC theory-to-data ratio is about 1.2 for \PDz mesons with $2<\pt<5\GeV$ and $-1<y<1$, lying at the upper edge of the measurement. For $\pt>5\GeV$, the ratio indicates that predictions exceed the data by a factor of 1.5--3. The measurement therefore provides scale-dependent constraints on calculations of charm quark production incorporating nonlinear QCD evolution.

To summarize, this Letter presented the first measurement of the inclusive, prompt and nonprompt, photonuclear \PDz meson cross section as a function of transverse momentum and rapidity in ultraperipheral heavy-ion collisions. Exploiting the clean environment of photonuclear interactions, where final state and hadronization effects are largely reduced compared to hadronic production, this measurement provides novel constraints on nuclear matter over a wide range of parton momentum fraction $x$ and interaction energy scale \Qsquared. Comparisons with theory provide discrimination among parton distribution function parametrizations and challenge calculations based on nonlinear quantum chromodynamics evolution at high \Qsquared. By demonstrating both its experimental feasibility and theoretical relevance, this study establishes open heavy-flavor production in ultraperipheral LHC collisions as a powerful probe of parton dynamics in nuclei.

\begin{acknowledgments}
We congratulate our colleagues in the CERN accelerator departments for the excellent performance of the LHC and thank the technical and administrative staffs at CERN and at other CMS institutes for their contributions to the success of the CMS effort. In addition, we gratefully acknowledge the computing centers and personnel of the Worldwide LHC Computing Grid and other centers for delivering so effectively the computing infrastructure essential to our analyses. Finally, we acknowledge the enduring support for the construction and operation of the LHC, the CMS detector, and the supporting computing infrastructure provided by the following funding agencies: SC (Armenia), BMBWF and FWF (Austria); FNRS and FWO (Belgium); CNPq, CAPES, FAPERJ, FAPERGS, and FAPESP (Brazil); MES and BNSF (Bulgaria); CERN; CAS, MoST, and NSFC (China); MINCIENCIAS (Colombia); MSES and CSF (Croatia); RIF (Cyprus); SENESCYT (Ecuador); ERC PRG, TARISTU24-TK10 and MoER TK202 (Estonia); Academy of Finland, MEC, and HIP (Finland); CEA and CNRS/IN2P3 (France); SRNSF (Georgia); BMFTR, DFG, and HGF (Germany); GSRI (Greece); NKFIH (Hungary); DAE and DST (India); IPM (Iran); SFI (Ireland); INFN (Italy); MSIT and NRF (Republic of Korea); MES (Latvia); LMTLT (Lithuania); MOE and UM (Malaysia); BUAP, CINVESTAV, CONACYT, LNS, SEP, and UASLP-FAI (Mexico); MOS (Montenegro); MBIE (New Zealand); PAEC (Pakistan); MES, NSC, and NAWA (Poland); FCT (Portugal);  MESTD (Serbia); MICIU/AEI and PCTI (Spain); MOSTR (Sri Lanka); Swiss Funding Agencies (Switzerland); MST (Taipei); MHESI and NSTDA (Thailand); TUBITAK and TENMAK (T\"{u}rkiye); NASU (Ukraine); STFC (United Kingdom); DOE and NSF (USA).
\end{acknowledgments}

\bibliography{auto_generated}
\cleardoublepage \appendix\section{The CMS Collaboration \label{app:collab}}\begin{sloppypar}\hyphenpenalty=5000\widowpenalty=500\clubpenalty=5000
\cmsinstitute{Yerevan Physics Institute, Yerevan, Armenia}
{\tolerance=6000
V.~Chekhovsky, A.~Hayrapetyan, V.~Makarenko\cmsorcid{0000-0002-8406-8605}, A.~Tumasyan\cmsAuthorMark{1}\cmsorcid{0009-0000-0684-6742}
\par}
\cmsinstitute{Institut f\"{u}r Hochenergiephysik, Vienna, Austria}
{\tolerance=6000
W.~Adam\cmsorcid{0000-0001-9099-4341}, J.W.~Andrejkovic, L.~Benato\cmsorcid{0000-0001-5135-7489}, T.~Bergauer\cmsorcid{0000-0002-5786-0293}, K.~Damanakis\cmsorcid{0000-0001-5389-2872}, M.~Dragicevic\cmsorcid{0000-0003-1967-6783}, C.~Giordano, P.S.~Hussain\cmsorcid{0000-0002-4825-5278}, M.~Jeitler\cmsAuthorMark{2}\cmsorcid{0000-0002-5141-9560}, N.~Krammer\cmsorcid{0000-0002-0548-0985}, A.~Li\cmsorcid{0000-0002-4547-116X}, D.~Liko\cmsorcid{0000-0002-3380-473X}, I.~Mikulec\cmsorcid{0000-0003-0385-2746}, J.~Schieck\cmsAuthorMark{2}\cmsorcid{0000-0002-1058-8093}, R.~Sch\"{o}fbeck\cmsAuthorMark{2}\cmsorcid{0000-0002-2332-8784}, D.~Schwarz\cmsorcid{0000-0002-3821-7331}, M.~Shooshtari, M.~Sonawane\cmsorcid{0000-0003-0510-7010}, W.~Waltenberger\cmsorcid{0000-0002-6215-7228}, C.-E.~Wulz\cmsAuthorMark{2}\cmsorcid{0000-0001-9226-5812}
% ...
\par}
\vskip\cmsinstskip
\dag:~Deceased\\
$^{1}$Also at Yerevan State University, Yerevan, Armenia\\
$^{2}$Also at TU Wien, Vienna, Austria\\
% ...
\end{sloppypar}
\end{document}